\documentclass[12pt,a4paper]{article}
\usepackage{cmap}
\usepackage[T2A]{fontenc}
\usepackage[cp1251]{inputenc}
\usepackage[russian,english]{babel}
\usepackage{amsmath,amsthm,amssymb}
\usepackage{amsopn}
\usepackage{bm} 
\usepackage{cite}
\usepackage[margin=2cm]{geometry} 

\newcommand{\R}{\mathbb{R}}  
\newcommand{\N}{\mathbb{N}}  
\newcommand{\Stable}{P_{\textup{stab}}} 
\newcommand{\Pack}{P_{\textup{pack}}} 
\newcommand{\Cover}{P_{\textup{cover}}} 
\newcommand{\Part}{P_{\textup{part}}} 
\newcommand{\DCP}{P_{\textup{2cover}}} 
\newcommand{\NPadj}{P_{\textup{matsui}}} 

\renewcommand{\emptyset}{\varnothing} 
\renewcommand{\le}{\leqslant}
\renewcommand{\ge}{\geqslant}
\DeclareMathOperator{\conv}{conv}
\DeclareMathOperator{\ext}{ext}
\newcommand{\lea}{\le_A} 

\newcommand{\symdiff}{\bigtriangleup}

\theoremstyle{plain}
\newtheorem{theorem}{Теорема}
\newtheorem{lemma}{Лемма}
\newtheorem{proposition}{Предложение}

\theoremstyle{definition}
\newtheorem{property}{Свойство}
\newtheorem{definition}{Определение}
\newtheorem{remark}{Замечание}

\usepackage{mathtools} 
\providecommand\given{} 
\newcommand\SetSymbol[1][]{%
	\nonscript\:#1\vert
	\allowbreak \nonscript\:	\mathopen{}}
\DeclarePairedDelimiterX\Set[1]\{\}{%
	\renewcommand\given{\SetSymbol[\delimsize]}	#1}

\usepackage{hyperref}

\newcommand{\keywords}[1]{\par\addvspace\baselineskip
\noindent{\bf Keywords:}\enspace\ignorespaces#1}

\begin{document}

\title{On the~family of 0/1-polytopes\\ with NP-complete non-adjacency relation%
\thanks{Supported by the~State Assignment for Research in P.G. Demidov Yaroslavl State University, 1.5768.2017/П220.}}
\author{A.N. Maksimenko}
\date{P.G. Demidov Yaroslavl State University, 150000, Yaroslavl, Sovetskaya 14\\
\url{maximenko.a.n@gmail.com}}

\maketitle


\begin{abstract}
In 1995 T.~Matsui considered a special family 0/1-polytopes 
for which the~problem of~recognizing the~non-adjacency of~two arbitrary vertices is NP-complete.
In~2012 the~author of~this paper established that all the~polytopes of~this family are present as faces in~the~polytopes associated with the~following NP-complete problems:
the~traveling salesman problem, the~3-satisfiability problem, the~knapsack problem, the~set covering problem, the~partial ordering problem, the~cube subgraph problem, and some others.
In~particular, it follows that for these families the~non-adjacency relation is also NP-complete.
On~the~other hand, it is known that the~vertex adjacency criterion is polynomial for polytopes of~the~following NP-complete problems:
the~maximum independent set problem, the~set packing and the~set partitioning problem, the~three-index assignment problem.
It is shown that none of~the~polytopes of~the~above-mentioned special family (with the~exception of~a one-dimensional segment) can be the~face of~polytopes associated with the~problems of~the~maximum independent set, of~a set packing and partitioning, and of~3-assignments.

\end{abstract}

\selectlanguage{russian}

Пусть $A\in\{0,1\}^{m\times n}$ "--- матрица инциденций элементов множества $G = \{g_1, \ldots, g_m\}$ и элементов некоторого множества $S = \{S_1, \ldots, S_n\} \subseteq 2^G$.
Выпуклая оболочка множества
\[
\Cover(A) = \Set*{\bm{x}\in\{0,1\}^n \given A \bm{x} \ge \bm{1}}
\]
называется \emph{многогранником покрытий}.
Семейство всех таких многогранников обозначаем $\Cover = \{\Cover(A)\}$.
Название многогранника объясняется тем, что каждый $\bm{x} \in \Cover(A)$ является характеристическим вектором некоторого набора множеств из $S$, покрывающих элементы множества $G$.
Так как $\Cover(A)$ содержит только 0/1-вектора, 
то оно совпадает с множеством вершин этого многогранника.
Далее, для краткости, мы часто будем называть многогранником множество его вершин. 

\emph{Многогранником двойных покрытий}~\cite{Maksimenko:2012} называется выпуклая оболочка множества
\begin{equation}
\DCP(B) = \Set*{\bm{x}\in\{0,1\}^n \given B \bm{x}  =  \bm{2}},
\label{eqDCP} 
\end{equation}
где $B$ "--- 0/1-матрица размера $m\times n$, каждая строка которой
содержит ровно четыре единицы, $\textbf{2}$ "--- $m$-мер\-ный вектор, 
все координаты которого равны 2.
Семейство всех таких многогранников обозначаем $\DCP = \{\DCP(B)\}$.

В 1995 году Т. Мацуи показал~\cite{Matsui:1995}, что задача распознавания несмежности вершин для некоторых многогранников двойных покрытий NP"=полна.
В~2012 году автор настоящей работы установил~\cite{Maksimenko:2012} (полная версия опубликована в~\cite{Maksimenko:2013}), что все многогранники семейства $\DCP$ присутствуют в качестве граней в многогранниках, ассоциированных со следующими NP"=полными задачами:
задача коммивояжера, задача 3"=выполнимость, задача о~рюкзаке, задача о~покрытии множества, задача о~частичном упорядочивании, задача о~кубическом подграфе и~некоторые другие.
Откуда, в~частности, следует, что для этих семейств задача распознавания несмежности вершин также NP-полна.
Тем самым, были обобщены результаты, полученные в работах Пападимитриу, Чунг, Гейст и Родин, Мацуи, Бондаренко и Юров, Альфаки и Мурти, Фиорини~\cite{Maksimenko:2013}.
С другой стороны, известно~\cite{Chvatal:1975,Ikura:1985,Balas:1989}, что критерий смежности вершин полиномиален для многогранников следующих NP-полных задач:
задача о независимом множестве в графе, задачи об упаковке и разбиении множества, трехиндексная задача о назначениях. 
В~\cite{Maksimenko:2016} было показано, что каждый многогранник из этих четырех семейств является гранью многогранника двойных покрытий, но, при этом, многогранник двойных покрытий
\[
 \conv\Set*{\bm{x}\in\{0,1\}^4 \given x_1 + x_2 + x_3 + x_4  =  2}
\]
не является гранью ни для одного многогранника из этих семейств.

Ниже мы рассмотрим подсемейство $\NPadj$ семейства $\DCP$, описанное Мацуи~\cite{Matsui:1995} и также имеющее NP-полный критерий несмежности вершин.
Мы покажем, что многогранники упомянутых выше четырех семейств с полиномиальным критерием смежности вершин являются гранями многогранников этого подсемейства.
С другой стороны, будет доказано, что \emph{ни один} из многогранников $\NPadj$, за исключением одномерного отрезка, не может быть гранью многогранников этих семейств.

%
%

\section{Определения и соглашения}
\label{sec:def}

Для $\{1,2,\dots,n\}$ будем пользоваться распространенным обозначением $[n]$.

В этой работе рассматриваются только \emph{0/1-многогранники}~--- выпуклые многогранники, вершины которых являются 0/1-векторами\footnote{Координаты 0/1-векторов принадлежат множеству $\{0,1\}$.}.
Более того, все многогранники определяются посредством описания множества их вершин (более подробно о способе описания~--- ниже) и рассматриваемые операции над многогранниками сводятся к операциям над множествами их вершин.
В частности, мы пользуемся следующими хорошо известными фактами.

Пусть $P$~--- многогранник, и $\alpha\colon P \to \R^k$~--- аффинное (линейное) отображение.
Тогда $Q = \alpha(P)$~--- многогранник, причем $\ext(Q) = \alpha(\ext(P))$, где $\ext(\cdot)$~--- множество вершин соответствующего многогранника.
В случае, если отображение $\alpha\colon P \to Q$ биективно, многогранники $P$ и $Q$ называются \emph{аффинно эквивалентными}.

Пусть $X$~--- множество вершин многогранника $P = \conv(X)$, а $H$~--- некоторая опорная\footnote{Гиперплоскость $H$ называется \emph{опорной} к многограннику $P$, если $P \cap H \neq \emptyset$ и многогранник $P$ лежит целиком с одной стороны от этой гиперплоскости.} гиперплоскость этого многогранника.
Множество $F = P \cap H$ называется \emph{гранью} многогранника $P$.
Нетрудно заметить, что $X \cap H$~--- множество вершин этой грани.

В настоящей работе мы не будем выходить за рамки этих двух операций (аффинное отображение и пересечение с опорной гиперплоскостью) над многогранниками.
Поэтому, в целях краткости, многогранник (а также его грани) отождествляется с множеством его вершин.

\begin{definition}\label{def:family}
	Каждое семейство многогранников, упоминаемое в настоящей работе, определяется тройкой:
	\begin{enumerate}
		\item \emph{Множество входных параметров} $S$, распознаваемое за полиномиальное время. (Например, для многогранника покрытий $\Cover(A)$ входными параметрами $s \in S$ является матрица $A \in \{0,1\}^{m\times n}$.)
		\item \emph{Размерность} $d = d(s)\in\N$, $s\in S$.
		\item \emph{Предикат допустимости} $g = g(s,x) \in \{\text{истина}, \text{ложь}\}$, $s\in S$, $x \in \{0,1\}^d$, вычисляемый за полиномиальное время.
	\end{enumerate}
	Каждый многогранник такого семейства однозначно определяется набором параметров $s\in S$, далее называемым \emph{кодом многогранника}, и представляет собой выпуклую оболочку множества
	\[
	X(s) = \Set*{x\in\{0,1\}^d \given d = d(s), \ g(s,x)}.
	\]
	Соответственно, $\Set{X(s)\given s\in S}$~--- все семейство многогранников.
\end{definition}

Заметим, что собственная размерность многогранника $\conv(X(s))$ может существенно отличаться от размерности $d(s)$ пространства, в котором он определен.

Для семейства многогранников из определения~\ref{def:family} задача проверки несмежности вершин формулируется следующим образом. Даны: код многогранника $s\in S$ и пара вершин $x_1, x_2 \in X(s)$. Верно ли, что $x$ и $y$ несмежны?
Данная задача принадлежит классу NP по следующим причинам.
Во-первых, условия $s\in S$ и $x_1, x_2 \in X(s)$ проверяются за полиномиальное время. 
Во-вторых, согласно теореме Каратеодори (см., например, \cite{Yemelichev:1981}), для несмежных вершин $x_1, x_2 \in X(s)$ обязательно найдется подмножество $Y \subseteq X(s) \setminus \{x_1, x_2\}$, состоящее не более, чем из $d$ вершин, и такое, что $\conv(Y) \cap \conv\{x_1, x_2\} \neq \emptyset$.

Заметим, что длина входа\footnote{Предполагается использование разумной схемы кодирования входных данных~\cite{Garey:1982}.} для задачи проверки несмежности вершин пропорциональна сумме длины кода $s$ и размерности $d(s)$.
В следующем определении эта сумма будет называться \emph{размером} соответствующего многогранника.

\begin{definition}
	\label{def:aff}
	Будем говорить, что семейство многогранников $P$ \emph{аффинно сводится} к семейству многогранников $Q$, если для каждого многогранника $p\in P$ найдется $q\in Q$ и аффинное отображение $\alpha \colon p \to q$
	такие, что
	\begin{enumerate}
		\item\label{item:1} Образ $\alpha(p)$ является гранью (возможно несобственной) многогранника $q$ и аффинно эквивалентен $p$.
		Далее этот факт обозначаем так: $p \lea q$.
		\item\label{item:2} Размерность пространства, в котором определен многогранник $q$, ограничена сверху полиномом от размерности пространства для $p$.
		\item\label{item:2s} Длина кода многогранника $q$ ограничена сверху полиномом от размера многогранника $p$.
		\item\label{item:3} Коэффициенты аффинного отображения $\alpha$ 
		вычисляются за полиномиальное время относительно размера многогранника $p$.
	\end{enumerate}
\end{definition}

\begin{remark}
	Приведенное выше определение отличается от определения аффинной сводимости в~\cite{Maksimenko:2013,Maksimenko:2016} наличием условий~\ref{item:2s} и~\ref{item:3} (полиномиальность длины кода многогранника $q$ и коэффициентов аффинного отображения).
	Тем не менее, для всех фактов аффинной сводимости, упоминаемых в работах~\cite{Maksimenko:2013,Maksimenko:2016}, справедливость этих условий легко проверяется, так как соответствующие аффинные отображения описаны явным образом.
\end{remark}

Как правило, доказательство аффинной сводимости семейства $P$ к семейству $Q$ выполняется по следующей схеме.
Для каждого $p\in P$ приводится описание многогранника $q\in Q$, его грани $F$ и биективного аффинного отображения $\alpha\colon p \to F$. 
Справедливость условий \ref{item:2}--\ref{item:3} определения \ref{def:aff}, как правило, очевидна.


Аффинная сводимость позволяет сравнивать различные характеристики семейств многогранников~\cite{Maksimenko:2016}.
В~частности, если семейство $P$ аффинно сводится к $Q$ и для многогранников из $P$ задача проверки несмежности вершин NP-полна, то для $Q$ эта задача также NP-полна.

По аналогии с многогранником покрытий определяются \emph{многогранник упаковок}
\[
\Pack(A) = \Set*{\bm{x}\in\{0,1\}^n \given A \bm{x} \le \bm{1}}
\]
и \emph{многогранник разбиений}~\cite[с.~135]{Yemelichev:1981}
\[
\Part(A) = \Set*{\bm{x}\in\{0,1\}^n \given A \bm{x} = \bm{1}},
\]
где матрица $A\in\{0,1\}^{m\times n}$ служит кодом многогранника.
Частным случаем многогранника упаковок является \emph{многогранник независимых множеств} графа $G = (V,E)$:
\[
\Stable(G) = \Set*{\bm{x}\in\{0,1\}^V \given x_v + x_u \le 1 \text{ для каждого ребра } \{v,u\} \in E}.
\]

Известно~\cite{Maksimenko:2016}, что семейства многогранников независимых множеств, многогранников упаковок, многогранников разбиений и многогранников трехиндексной задачи о назначениях аффинно сводятся друг к другу.

%
%

\section{Многогранники Мацуи}

Далее, с целью упрощения рассуждений, мы рассмотрим более удобное, чем в первоисточнике~\cite{Matsui:1995} описание специального подсемейства $\NPadj$ семейства многогранников двойных покрытий.

Прежде всего отметим, что вопрос <<Содержит ли многогранник разбиений $\Part(A)$ хотя бы одну точку?>> является NP-полной задачей даже если каждая строка матрицы $A$ содержит ровно три единицы~\cite{Matsui:1995,Garey:1982}.
С каждым многогранником $\Part(A)$, где $A \in \{0,1\}^{m\times n}$ содержит ровно три единицы в каждой строке, свяжем многогранник, множество вершин $\NPadj(A) \subset \{0,1\}^{3n+3}$ которого определим следующим образом.
Для трех координат вектора $\bm{x} \in \NPadj(A)$ введем особые обозначения: $y_1, y_2, y_3$. 
Каждой координате $z_j$, $j\in [n]$, вектора $\bm{z} = (z_1,\dots,z_n) \in \Part(A)$ будут соответствовать три координаты $x_j$, $\bar{x}_j$ и $x'_j$ вектора $\bm{x} \in \NPadj(A)$, и два ограничения
\begin{align}
x_j + \bar{x}_j &= 1, \label{eq:adj01}\\
y_1 + y_2 + x'_j + \bar{x}_j&= 2. \label{eq:adj2}
\end{align}
А для каждого ограничения вида $z_i + z_j + z_k = 1$ из описания $\Part(A)$
(случай, когда $A$ не содержит ни одной строки, исключаем из рассмотрения)
добавим к описанию множества $\NPadj(A)$ уравнение
\begin{equation}
y_3 + x_i + x'_j + x'_k = 2. \label{eq:adj3}
\end{equation}
Итак, $\NPadj(A)$ "--- множество 0/1-векторов из $\{0,1\}^{3n+3}$, удовлетворяющих ограничениям \eqref{eq:adj01}--\eqref{eq:adj3}.

Заметим, что каждое уравнение~\eqref{eq:adj01} в описании многогранника $\NPadj(A)$ можно заменить на 
\[
a + b + x_j + \bar{x}_j = 2, \quad a = 0, \quad b = 1.
\]
Следовательно, для каждой матрицы $A \in \{0,1\}^{m\times n}$, имеющей ровно три единицы в каждой строке, несложно описать матрицу $B \in \{0,1\}^{(2n+m)\times(3n+5)}$,
что $\NPadj(A)$ является гранью многогранника $\DCP(B)$, 
лежащую в пересечении (опорных для $\DCP(B)$) гиперплоскостей $a=0$ и $b=1$. 
Более того, в~\cite{Matsui:1995} описан многогранник двойных покрытий, аффинно эквивалентный $\NPadj(A)$, но его описание требует большего числа переменных и уравнений и поэтому здесь не рассматривается.

Обратим теперь внимание на то, что ограничения 
\[
y_1 = 0, \quad y_2 = 1, \quad y_3 = 1
\]
определяют грань многогранника $\NPadj(A)$, аффинно эквивалентную многограннику $\Part(A)$.
Таким образом,
\begin{equation}\label{eq:PartNPadj}
\Part(A) \lea \NPadj(A),
\end{equation}	
при условии, что в каждой строке матрицы $A$ содержится ровно три единицы.
То же самое верно и для следующих наборов ограничений:
\begin{enumerate}
	\item[1)] $y_1 = 1$, $y_2 = 0$, $y_3 = 1$;
	\item[2)] $y_1 = 0$, $y_2 = 1$, $y_3 = 0$;
	\item[3)] $y_1 = 1$, $y_2 = 0$, $y_3 = 0$.
\end{enumerate}
Введем для этих граней (точнее, множеств их вершин) следующие обозначения:
\begin{align*}
F_1 &= \Set*{\bm{x}\in \NPadj(A) \given y_1 = 0, \ y_2 = 1, \ y_3 = 1}, \\
F_2 &= \Set*{\bm{x}\in \NPadj(A) \given y_1 = 1, \ y_2 = 0, \ y_3 = 1}, \\
F_3 &= \Set*{\bm{x}\in \NPadj(A) \given y_1 = 0, \ y_2 = 1, \ y_3 = 0}, \\
F_4 &= \Set*{\bm{x}\in \NPadj(A) \given y_1 = 1, \ y_2 = 0, \ y_3 = 0}.
\end{align*}
Заметим, что никакие две из этих четырех граней не имеют общих точек.
Кроме того,
\begin{equation}
\label{eq:F4F3}
F_4 = \Set*{\bm{1} - \bm{x} \given \bm{x} \in F_1}, \qquad
F_3 = \Set*{\bm{1} - \bm{x} \given \bm{x} \in F_2}.
\end{equation}

Ограничениям
\[
y_1 = y_2 = 0
\]
удовлетворяет ровно одна вершина многогранника $\NPadj(A)$, имеющая координаты
\[
x'_j = \bar{x}_j = 1, \quad y_3 = x_j = 0, \qquad j\in[n].
\]
Обозначим эту вершину $\bm{x^0}$.
Аналогично, если 
\[
y_1 = y_2 = 1,
\]
то 
\[
x'_j = \bar{x}_j = 0, \quad y_3 = x_j = 1, \qquad j\in[n].
\]
Обозначим эту вершину $\bm{\bar{x}^0}$.
Очевидно, $\bm{x^0} + \bm{\bar{x}^0} = \bm{1}$.

Из приведенных выше рассуждений следует, что 
\[
\NPadj(A) = F_1 \cup F_2 \cup F_3 \cup F_4 \cup \{\bm{x^0}, \bm{\bar{x}^0}\},
\]
причем никакие два из этих пяти множеств не имеют общих точек.
Кроме того, вершины $\bm{x^0}$ и $\bm{\bar{x}^0}$ смежны тогда и только тогда, когда $F_1 = \emptyset$ (в противном случае $\conv\{F_1 \cup F_4\}$ и $\conv\{\bm{x^0}, \bm{\bar{x}^0}\}$ имеют общую точку $\bm{1}/2$).
Таким образом, в силу того, что $F_1$ аффинно эквивалентна $\Part(A)$, а проверка неравенства $\Part(A) \ne \emptyset$ NP"=полна, приходим к следующему выводу.

\begin{theorem}[Мацуи {\cite[theorem 4.1]{Matsui:1995}}]
Задача проверки несмежности вершин $\bm{x^0}$ и $\bm{\bar{x}^0}$ многогранника $\NPadj(A)$ NP-полна.
\end{theorem}

Покажем теперь, что соотношение~\eqref{eq:PartNPadj} выполнено и в тех случаях, когда число единиц в строках матрицы $A$ отличается от трех.
Точнее, для каждой матрицы $B$ (с любым числом единиц в строках) можно построить матрицу $A$, имеющую три единицы в каждой строке, что
\[
\Part(B) \lea \NPadj(A).
\]

Так как семейство $\Part$ аффинно сводится к $\Stable$~\cite{Maksimenko:2016}, достаточно показать, что для любого графа $G=(V,E)$, $|V|>0$, $|E|>0$, можно построить матрицу $A$ с тремя единицами в каждой строке, что $\Stable(G) \lea \Part(A)$.

Заметим, что неравенство $x_v + x_u \le 1$, при условии $x_v, x_u \in \{0,1\}$, из определения многогранника $\Stable(G)$ можно заменить равенством $x_v + x_u + x_{vu} = 1$, где $x_{vu} \in \{0,1\}$ "--- вспомогательная переменная, линейно зависимая от $x_v$ и $x_u$. 
Следовательно, многогранник $\Stable(G)$, $G=(V,E)$, аффинно эквивалентен некоторому $\Part(A)$, где матрица $A \in \{0,1\}^{|E|\times(|V|+|E|)}$ содержит ровно три единицы в каждой строке~\cite{Maksimenko:2016}.
Таким образом, из соотношения~\eqref{eq:PartNPadj} следует

\begin{proposition}
	Семейства $\Stable$, $\Part$, $\Pack$ и многогранники трехиндексной задачи о назначениях аффинно сводятся к $\NPadj$.
\end{proposition}

Как известно~\cite{Chvatal:1975}, многогранники $\Stable(G)$ имеют простой критерий проверки смежности вершин.
Соответственно, в предположении $\textup{NP} \ne \textup{P}$ аффинная сводимость $\NPadj$ к $\Stable$ невозможна.
Покажем, 
что и без условия $\textup{NP} \ne \textup{P}$ ни один многогранник семейства $\NPadj$ не может быть гранью многогранников семейства $\Stable$.

\begin{theorem}\label{thm:DCPStable}
	Если многогранник $\NPadj(A)$ не является отрезком, то 
	$\NPadj(A) \lea \Stable(G)$ невозможно ни для какого графа $G$.
\end{theorem}

%
%

\section[Доказательство теоремы]{Доказательство теоремы \ref{thm:DCPStable}}

Как было замечено выше, многогранник $\NPadj(A)$ обязательно содержит пару вершин $\bm{x^0}$ и $\bm{\bar{x}^0}$, и некоторое количество 
четверок вершин вида $\bm{x^{2i-1}} \in F_1$, $\bm{\bar{x}^{2i-1}} \in F_4$, $\bm{x^{2i}} \in F_2$, $\bm{\bar{x}^{2i}} \in F_3$, $i\in [k]$, $k \ge 1$.
Причем, согласно~ \eqref{eq:F4F3},
\begin{equation}
\label{3xpairs}
\bm{x^0} + \bm{\bar{x}^0} = \bm{x^{2i-1}} + \bm{\bar{x}^{2i-1}} = \bm{x^{2i}} + \bm{\bar{x}^{2i}}.
\end{equation}

Предположим, что $\NPadj(A)$ аффинно эквивалентен некоторой грани 
\[
H = \{\bm{y^0}, \bm{\bar{y}^0}, \ldots, \bm{y^{2k}}, \bm{\bar{y}^{2k}}\}
\] 
многогранника $\Stable(G)$ для некоторого графа $G = (V,E)$.
Очевидно, вершины этой грани должны наследовать свойство~\eqref{3xpairs}:
\begin{equation}
\label{3pairs}
\bm{y^0} + \bm{\bar{y}^0} = \bm{y^{2i-1}} + \bm{\bar{y}^{2i-1}} = \bm{y^{2i}} + \bm{\bar{y}^{2i}}.
\end{equation}
Покажем, что в многограннике $\Stable(G)$ есть еще пара вершин $\bm{y^*}$ и $\bm{\bar{y}^*}$, для которых
\begin{equation}
\label{ThGoal}
\bm{y^*} + \bm{\bar{y}^*} = \bm{y^0} + \bm{\bar{y}^0}.
\end{equation}
Это будет означать, что пересечение $\conv\{\bm{y^*}, \bm{\bar{y}^*}\}$ и $\conv(H)$ не пусто.
То есть $H$ не является гранью $\Stable(G)$.

\begin{figure}[hb]
	\[
	\begin{aligned}
	\bm{y^0}&=(\overbrace{\text{\texttt{1,0,1,1,}}}^{I}
	\overbrace{\text{\texttt{0,0,0,0,1,1,1}}}^{J})\\[-1.0ex]
	&\phantom{=(\text{\texttt{1,0,1,1, }}}^{j_0} \\[-2.0ex]
	\bm{\bar{y}^0}&=(\text{\texttt{1,0,1,1,}}
	\underbrace{\text{\texttt{1,1,1,1,}}}_{U_0\vphantom{\bar{U}_0}}\!
	\underbrace{\text{\texttt{0,0,0}}}_{\bar{U}_0})
	\end{aligned}
	\]
	\caption{Множества индексов $I$, $J$, $U_0$, $\bar{U}_0$.}
	\label{fig:IJU}
\end{figure}

Пусть $\bm{y^0} = (y^0_1, \ldots, y^0_m)$ и $\bm{\bar{y}^0} = (\bar{y}^0_1, \ldots, \bar{y}^0_m)$, где $m$ "--- число вершин графа $G$.
Рассмотрим множество
\[
I = \Set*{i\in [m] \given y^0_i = \bar{y}^0_i}.
\]
Так как каждая вершина в $\Stable(G)$ является 0/1-вектором, то из~\eqref{3pairs} и~\eqref{ThGoal} следует
\begin{equation}
\label{eq:8}
y^*_i = \bar{y}^*_i = y^0_i = \bar{y}^0_i = \cdots = y^{2k}_i = \bar{y}^{2k}_i  \quad \text{ при } i\in I.
\end{equation}
Далее будем рассматривать только те координаты, значения которых различны для каждой пары вершин (см. рис.~\ref{fig:IJU}):
\[
J = \Set*{j\in [m] \given y^0_j + \bar{y}^0_j = 1} = [m] \setminus I.
\]
Очевидно, $J \ne \emptyset$.

Зафиксируем какой нибудь индекс $j_0 \in J$ и для каждого $i \in \{0,1,\dots, 2k\}$ определим множество
\[
U_i =\begin{cases}
\Set*{j\in J \given y^i_j = 1},& \text{если } y^i_{j_0} = 1,\\
\Set*{j\in J \given \bar{y}^i_j = 1},& \text{иначе.}
\end{cases} 
\]
По построению все эти множества попарно различны и $j_0 \in U_i$ (см. рис.~\ref{fig:IJU}).
Для каждого $U_i$ рассмотрим его дополнение $\bar{U}_i = J \setminus U_i$.
Непосредственно из описания $U_i$ и $\bar{U}_i$ следует

\begin{property}\label{prop:1}
Для любого $i \in \{0,1,\dots, 2k\}$ и для любых $p,r \in U_i$ (а также для любых $p,r \in \bar{U}_i$) найдется вершина $\bm{y} \in H$ такая, что $y_p = y_r = 1$. 
То есть неравенство $y_p + y_r \le 1$ отсутствует в описании многогранника $\Stable(G)$.
\end{property}

Далее нам понадобится определение \emph{симметрической разности} двух множеств $X$ и $Y$:
\[
X \symdiff Y = (Y \setminus X) \cup (X \setminus Y).
\]
Симметрическая разность обладает следующими свойствами:
\begin{enumerate}
	\item $X \symdiff Y = \emptyset \iff X = Y$.
	\item Результат выражения $X \symdiff Y \symdiff Z$ не зависит от перестановки множеств и порядка выполнения операций.
	\item $X \symdiff Y = Z \iff X \symdiff Z = Y$.
\end{enumerate}

Введем в рассмотрение множество
\[
S = S(i,j,t) = U_i \symdiff U_j \symdiff U_t, \quad 0\le i < j < t \le 2k,
\]
Рассмотрим вектор $\bm{y^*} = \bm{y^*}(S)$ с координатами
\[
y^*_i = \begin{cases}
y^0_i,& \text{при } i \in I,\\
1,& \text{при } i \in S,\\
0,& \text{при } i \in J \setminus S,
\end{cases}
\]
и вектор $\bm{\bar{y}^*} = \bm{\bar{y}^*}(S)$ с координатами
\[
\bar{y}^*_i = \begin{cases}
y^0_i,& \text{при } i \in I,\\
0,& \text{при } i \in S,\\
1,& \text{при } i \in J \setminus S,
\end{cases}
\]

\begin{lemma}
Векторы $\bm{y^*}$ и $\bm{\bar{y}^*}$ принадлежат $\Stable(G)$.
\end{lemma}
\begin{proof}
Достаточно показать, что если координаты вершин грани $H$ удовлетворяют некоторому неравенству вида $y_p + y_r \le 1$ из описания многогранника $\Stable(G)$, то координаты точек $\bm{y^*}$ и $\bm{\bar{y}^*}$ тоже ему удовлетворяют.

Возможны несколько случаев.

\textbf{I.} 
Пусть $p,r\in I$, $p \ne r$.
Так как при $i \in I$ $i$-е координаты вершин грани $H$ и векторов $\bm{y^*}$ и $\bm{\bar{y}^*}$ совпадают, то из того, что неравенство $y_p + y_r \le 1$  выполнено для $H$ следует, что оно также выполнено и для векторов $\bm{y^*}$ и $\bm{\bar{y^*}}$.

\textbf{II.} 
Пусть $p \in I$, $r \in J$.
(Случай $r \in I$, $p \in J$ разбирается аналогично.)
Тогда $y^0_r + \bar{y}^0_r = y^*_r + \bar{y}^*_r = 1$.
Следовательно, $\max\{y^*_r, \bar{y}^*_r\} = \max\{y^0_r, \bar{y}^0_r\} = 1$.
И опять из выполнения неравенства $y_p + y_r \le 1$ для $H$ следует, что оно также выполнено для $\bm{y^*}$ и $\bm{\bar{y^*}}$.

\textbf{III.} 
Пусть $p \in S$, $r \in J\setminus S$.
(Случай $r \in S$, $p \in J\setminus S$ разбирается аналогично.)
Тогда $y^*_p + y^*_r = \bar{y}^*_p + \bar{y}^*_r = 1$ и требуемое ограничение выполнено.

\textbf{IV.} 
Пусть $p,r \in S$, $p \ne r$, где $S = S(i,j,t)$.
(Случай $p,r \in J \setminus S$, $p \ne r$, разбирается аналогично.)
Покажем, что в этом случае $p$ и $r$ принадлежат одновременно одному из шести множеств:
$U_i$, $U_j$, $U_t$, $\bar{U}_i$, $\bar{U}_j$, $\bar{U}_t$.
Если это действительно так, то, согласно свойству~\ref{prop:1}, неравенство $y_p + y_r \le 1$ отсутствует в описании многогранника $\Stable(G)$.

Итак, предположим, что $p,r \in U_i \symdiff U_j \symdiff U_t$, и покажем, что тогда $p$ и $r$ принадлежат одновременно одному из множеств:
$U_i$, $U_j$, $U_t$, $\bar{U}_i$, $\bar{U}_j$, $\bar{U}_t$.
Заметим, что $U_i \symdiff U_j \symdiff U_t$ представляет собой объединение четырех множеств:
\[
U_i \symdiff U_j \symdiff U_t = (U_i \cap U_j \cap U_t) \cup (U_i \cap \bar{U}_j \cap \bar{U}_t) \cup (\bar{U}_i \cap U_j \cap \bar{U}_t) \cup (\bar{U}_i \cap \bar{U}_j \cap U_t).
\]
Если $p$ и $r$ принадлежат одному из этих четырех множеств, то требуемое условие выполнено.
Нетрудно проверяется, что условие выполнено и в случае, когда $p$ и $r$ принадлежат разным множествам.
Например, если $p \in U_i \cap \bar{U}_j \cap \bar{U}_t$, а $r \in \bar{U}_i \cap \bar{U}_j \cap U_t$, то $p,r \in \bar{U}_j$.
\end{proof}

Для завершения доказательства теоремы остается показать, что найдется множество $S$ такое, что вектор $\bm{y^*} = \bm{y^*}(S)$ (а вместе с ним и вектор $\bm{\bar{y}^*} = \bm{\bar{y}^*}(S)$) будет отличаться от всех остальных вершин грани $H$.

\begin{lemma}
Существует $t \in \{2, 3, \dots, 2k\}$ такое, что $S(0,1,t)$ отличается от каждого из множеств $U_p$ и $\bar{U}_p$, $0\le p \le 2k$.
\end{lemma}
\begin{proof}
Начнем с простого случая, когда $k=1$.
Нетрудно проверить, что
\[
S(0,1,2) = U_0 \symdiff U_1 \symdiff U_2 \not\in \{U_0, U_1, U_2, \bar{U}_0, \bar{U}_1, \bar{U}_2\},
\]
так как все множества различны и, кроме того, все $U_i$ имеют общий элемент $j_0$:
\[
U_i \symdiff U_j \ne \emptyset \quad \text{и} \quad U_i \symdiff U_j \ne J, \quad \text{при } i\ne j.
\]

Предположим теперь, что $k>1$.
Рассмотрим тройки вида 
\[
U_0 \symdiff U_1 \symdiff U_i, \quad 2 \le i \le 2k.
\]
Как было замечено выше,
\[
U_0 \symdiff U_1 \symdiff U_i \not\in \{U_0, U_1, U_i\}.
\]
Кроме того, $U_0 \symdiff U_1 \symdiff U_i \ne \bar{U}_j$ при любом $j \in \{0,1,\dots,2k\}$, так как $j_0 \notin \bar{U}_j$.
Предположим, что 
\begin{equation}
\label{eq:sym-pair}
U_0 \symdiff U_1 \symdiff U_i = U_j
\end{equation}
при некотором $j \in \{2,3,\dots,2k\} \setminus \{i\}$.
Но тогда, в силу свойств симметрической разности,
\[
U_0 \symdiff U_1 \symdiff U_j = U_i.
\]
Причем
\[
U_0 \symdiff U_1 \symdiff U_t \ne U_j, \quad \forall t \ne i,
\]
так как иначе $U_t = U_i$, что невозможно по условию.
По тем же соображениям,
\[
U_0 \symdiff U_1 \symdiff U_t \ne U_i, \quad \forall t \ne j.
\]
Таким образом, все возможные индексы $i$ и $j$, для которых выполнено условие \eqref{eq:sym-pair}, разбиваются на непересекающиеся пары. 
Но множество $\{2,3,\dots,2k\}$ содержит нечетное число индексов.
Значит, обязательно найдется $i \in \{2,3,\dots,2k\}$, для которого
$S(0,1,i) = U_0 \symdiff U_1 \symdiff U_i$ будет отличаться от каждого из множеств $U_p$ и $\bar{U}_p$, $0\le p \le 2k$.
\end{proof}

\section{Благодарности}

Настоящая работа появилась благодаря вопросу М.\,Н.~Вялого, заданному на одном из семинаров в Вычислительном центре РАН.
Особых слов благодарности заслуживает анонимный рецензент, ценные замечания которого привели к появлению раздела~\ref{sec:def}.
 
%
%

\end{document}